\documentclass[useAMS,usenatbib,usegraphicx]{mn2e}

\newcommand{\vt}{\mbox{\bf {T}}}

\newcommand{\vm}{\mbox{\bf {M}}}
\newcommand{\vn}{\mbox{\bf {N}}}
\newcommand{\vf}{\mbox{\bf {F}}}
 
\input epsf
\def\plotone#1{\centering \leavevmode       
\epsfxsize=\columnwidth \epsfbox{#1}}         

\title{On the Presence of Thermal SZ Induced Signal in the First Year WMAP
Temperature Maps}

\author[C. Hern\'andez--Monteagudo \& J.A. Rubi\~no-Mart\1n]{C.
Hern\'andez--Monteagudo$^{1}$\thanks{E-mail:chm@MPA-Garching.MPG.DE} and
J.A. Rubi\~no--Mart\1n$^{1}$\\
$^{1}$Max-Planck Institut f\"ur Astrophysik, Karl-Schwarzschild-Str. 1,
D-85740 Garching, Germany\\
}
\begin{document}

\date{}

\pagerange{\pageref{firstpage}--\pageref{lastpage}} \pubyear{2003}

\maketitle

\label{firstpage}

\begin{abstract}

Using available optical and X-ray  catalogues of clusters and superclusters
of galaxies, we build templates of tSZ emission as they should be detected
by the WMAP experiment. We compute the cross-correlation of our templates
with WMAP temperature maps, and interpret our results separately for clusters
and for superclusters of galaxies. For clusters of galaxies, we claim 
2-5 $\sigma$ detections in our templates built from BCS 
\citep{ebeling98,ebeling00}, NORAS \citep{bohringer00} and 
\citet{grandi99} catalogues. 
In these templates, the typical cluster temperature
decrements in WMAP maps are around 15-35 $\mu$K in the RJ range 
(no beam deconvolution applied). 
Several tests probing the possible influence of foregrounds in our analyses
demonstrate that our results are robust against galactic contamination.
On supercluster scales, we detect a diffuse component in the V \& W WMAP
bands which cannot be generated by superclusters in our catalogues 
\citep{einasto94,einasto97}, and
which is not present in the clean map of Tegmark, de Oliveira-Costa \&
Hamilton (2003). 
Using this clean map, our analyses yield, for Einasto's supercluster 
catalogues, the following upper limit for the
comptonization parameter associated to supercluster scales: $y_{SC} < 2.18 \times 10^{-8}$ at the 95\% confidence limit. 

\end{abstract}

\begin{keywords}
cosmic microwave background -- galaxies: clusters: general -- diffuse radiation -- intergalactic medium
\end{keywords}

\section{Introduction}

The WMAP experiment is currently observing the microwave sky at five different
frequency channels with different angular resolution \citep{Bennett03a}.
 The low 
frequency bands, K, Ka \& Q, measuring at 23, 33 \& 41GHz respectively,
are expected to be particularly sensitive to the free-free and synchrotron
emission in the Milky Way. Although their angular resolution is not as good as
in the high
frequency channels, their measurements of the foreground contamination
are critical in order to achieve an optimal subtraction of non-cosmological
signal in the overall analysis. This analysis has yielded the identification
of two main components in the sky signal: a cosmological component, which
constrains crucial cosmological parameters such as 
 $\Omega_{\Lambda}$, $\Omega_{m}$, $n$, $\tau_{reion}$ with unprecedented
accuracy \citep{Bennett03b}, and a foreground-induced component, whose
impact in the high frequency channels (V (61GHz) \& W(94GHz)) is modelled
by means of the low frequency measurements. These channels
have been located in a frequency range where the contribution from 
foregrounds is expected to be minimal \citep{Bennett03c}, and their
high angular resolution (up to $0.21\degr$) enables the study the sub-horizon
structure of the Last Scattering Surface.

However, it is also expected that the presence of Large Scale Structure
intersecting the geodesics of the CMB photons leave a signature in the form
of {\it secondary} temperature anisotropies. Among these temperature 
fluctuations, the most important (in terms of amplitude) are those caused by 
the Integrated Sachs-Wolfe effect, (ISW), due to the variation of the
gravitational potential, and the thermal Sunyaev-Zel'dovich
(hereafter tSZ, \citet{sunyaev80}), due to the distortion of the
black body spectrum of CMB photons after interacting with a hot electron 
plasma. The search of tSZ-induced signal in CMB data sets by comparing them 
with surveys of Large Scale Structure
has been performed for COBE data \citep{ganga93,Boughn93,banday96,kneissl97},
 and Tenerife data \citep{rubino00}. They all failed to measure any significant
correlation, and hence could only set upper limits to the comptonization
parameter $y$.

In this context, three recent analyses have been carried out on WMAP data.
This first one has been carried out by the WMAP team: 
a cross-correlation of the W-band with the XBACs catalog of 242 
Abell-type clusters \citep{ebeling96} has prompted a detection at 
2.5$\sigma$ level. However, a direct cross-correlation of WMAP data with ROSAT 
\citep{diego03} has given negative results. In this analysis,
the difference map between V and W bands was compared to the ROSAT map by the
study of the power spectrum of the product map and the cross-power spectrum
of both maps. Apart from an apparently spurious coincidence in the product
map, there is no trace for any significant correlation. More recently,
\citet{Boughn03} have reported a positive cross-correlation at the
2--3 sigma level between the WMAP data and the X-ray HEAO-1 and radio 
NRAO VLA sky surveys. Their study focuses on extra-galactic objects 
as tracers of
the Large Scale Structure, and the positive sign of their correlation would
point to the ISW effect as the responsible for this excess. A similar result
was reported by \citet{pablo03} after cross-correlating the V band of WMAP
with a template built from the APM galaxy survey. They find a positive 
cross correlation with a significance of $\sim 2\sigma$.

In this work, we extend the study of possible correlation of WMAP
data with Large Scale Structure (LSS) to cluster and supercluster scales. 
By using cluster surveys present in the literature, we construct template maps 
of LSS, and cross correlate them with the temperature
maps of WMAP. Although the angular resolution and the sensitivity of WMAP
are not ideal for the typical tSZ amplitudes and angular profiles of clusters 
of galaxies, we expect that if a large enough number of them are indeed 
contributing to the map their final imprint should be statistically measurable.
 On the other hand, angular sizes of superclusters of galaxies are 
resolved by the V \& W bands of WMAP, so they should allow to set constraints
in the density and/or temperature of the diffuse gas thought to be present
 in those structures.

In Section 2 we outline the details of the statistical methods used in
our analyses. Section 3 describes the WMAP data, 
whereas in Section 4 we explain how our tSZ templates of clusters and 
supercluster of galaxies were synthesised. Finally, in Section 5 we show 
our results, which we discuss in Section 6.

\section{Statistical Method}

In this section we describe the statistical tools used to search for
signature of Large Scale Structure in WMAP CMB maps in the form of 
tSZ effect. All our statistics will
be defined in the real space, and will be restricted to those regions where
we expect a high contribution from SZ sources. By doing this, we 
are trying  to minimise
the effects related to the limited angular resolution
and sensitivity of WMAP in terms of typical SZ amplitudes. 
A Fourier analysis
in a selected and non-continuous sample of pixels cannot always be 
defined in a simple way and will be discarded here.
Core radii of clusters of 
galaxies usually
subtend an angle of $\simeq 1'-3'$ in the sky, which is considerably
smaller than the FWHM of the highest resolution channel (94GHz) of WMAP,
 ($\sim 12.6$ arcmins).
For this reason, from the five bands of WMAP, we shall only use this channel
 when searching for cluster induced signal in the WMAP data, 
as we expect that other channels
will dilute the SZ signal to a much greater extent. The high resolution map
of  Tegmark, de Oliveira-Costa \& Hamilton (2003), which combines all bands
but preserves the angular resolution of the 94GHz channel,
 will also be considered.
However, for superclusters of
galaxies, the situation differs considerably, provided the fact that
 these sources can have a size of a few
degrees on the sky, enabling the use of lower frequency channels of WMAP.\\

The two statistical methods used in this paper will search for similarities
present in both our templates and the WMAP CMB maps.
The first method (method {\it i}) consists of a pixel-to-pixel comparison of 
CMB maps to the
templates. Given the CMB and noise amplitudes in those pixels, we estimate
what is the fraction of signal present in CMB maps which is correlated to
the structure of our catalogues.
 In the second method, (method {\it ii}) 
 we computed the cross-correlation function
between our templates and the WMAP map(s), and compared it to the 
auto-correlation function computed in our synthesised maps.

We shall assume that the WMAP data can be modelled as the following
sum of components:
\begin{equation}
\vt = \vt_{cmb} + \alpha \vm + \sum_i c_i \vf^{i} + \vn,
\label{eq:Ttot}
\end{equation}
where $\vt_{cmb}$ denotes the CMB, $\vn$ the noise, and 
$\vm$ is the SZ induced signal
modelled by our templates, which enters into the map with an average
amplitude given by $\alpha$. $c_i \vf^{i}$ denotes the {\it ith} foreground
component. This last term will introduce a bias in our statistics, and its
impact in our results must be considered. 
There might be as well some offset due to the inaccurate setting of the
zero level in the WMAP maps. From figure (7) in \citet{Bennett03c} one
would expect a typical error of a few ($\sim 3-5$) microkelvin. The impact
of such offsets will be addressed when quoting our results.
 \\

{\bf Method {\it i)}:}\\
We shall minimise the statistic $\chi^2$ given by
 
\begin{equation}
\chi^2 = (\vt - \alpha \vm) {\cal C}^{-1} (\vt - \alpha \vm)^T
\label{eq:chi1}
\end{equation}

\noindent in terms of $\alpha$. By doing this, our estimate  of 
$\alpha$ will be equal to:
\begin{equation}
E[\alpha] = \frac{ \vt {\cal C}^{-1} \vm^T} { \vm {\cal C}^{-1} \vm^T }
\label{eq:alpha1}
\end{equation}
In both equations \ref{eq:chi1} and \ref{eq:alpha1}, ${\cal C}$ denotes
the correlation function of the SZ-free temperature map, i.e.:
\begin{equation}
{\cal C} = < (\vt - \alpha \vm)^T (\vt - \alpha \vm) > 
\label{eq:C1}
\end{equation}
In our analyses, we considered only CMB and noise when computing ${\cal C}$.
The cosmological signal was modelled by using the CMB power spectrum
measured by the WMAP team \citep{Bennett03a}, whereas the noise component
was introduced using the amplitudes at each pixel according to the scanning
strategy of the WMAP mission\footnote{All data related specifically to
the WMAP mission has been obtained from: http://lambda.gsfc.nasa.gov}. 
The formal error for the estimated $\alpha$ is given by:
\begin{equation}
\sigma_\alpha =  \sqrt{ \frac{1} { \vm {\cal C}^{-1} \vm^T } }
\label{eq:sgalpha1}
\end{equation}

However, in these equations we have neglected the bias term introduced by
the presence of any possible foreground component, (i.e. synchrotron, 
free-free, dust or point sources).  
Our basic assumption will be that foregrounds and tSZ sources will
introduce temperature fluctuations of opposite sign, 
positive and negative respectively, in the WMAP CMB scans. We remark 
at this point 
that all WMAP channels are in the Rayleigh-Jeans (RJ) frequency range,
for which the tSZ distortions of the CMB planckian spectrum introduce
{\it negative} temperature fluctuations.
Therefore, because of the presence of
foreground residuals in WMAP maps, the measured $\alpha$ will
be shifted to positive values, and we will only be able to establish
lower limits to the presence of SZ-induced signal in WMAP scans, so our
approach can be regarded as conservative. \\

{\bf Method {\it ii)}: }\\
We shall define the $\chi^2$ statistic in terms of the
cross-correlation and auto-correlation functions computed from the
WMAP CMB map(s) and our templates. In this case, we write $\chi^2$ in the form:

\[
 \chi^2 =  \sum_{ij} \left[ C_{TM}(\theta_i) - 
	\alpha C_{MM}(\theta_i) \right]  
\]
\begin{equation}
	\phantom{xxxxxxxxxxxxxxxxxxx}	  {\mbox{\bf $\pi$}}^{-1}_{ij}
	\left[ C_{TM}(\theta_j) - \alpha C_{MM}(\theta_j) \right]
\label{eq:chi2b}
\end{equation}

In this equation, $C_{TM}$ and $C_{MM}$ denote the cross and auto correlation
function of our template maps with the WMAP scans, respectively. 
${\mbox{\bf $\pi$}}$ is the covariance matrix, defined as:
\[
{\mbox{\bf $\pi$}}_{ij} =
\]
\begin{equation}
 \langle \;\left(
	 C_{TM}(\theta_i) - \langle  C_{TM}(\theta_i)
						\;\rangle\; \right)\;
				\left(
	 C_{TM}(\theta_j) - \langle  C_{TM}(\theta_j)
						\;\rangle\; \right) \;
			\rangle .
\label{eq:pi} 
\end{equation}

The angle brackets denote ensemble averages after computing the
cross-correlation between our templates and 100 Monte-Carlo realizations
of the microwave sky accounting for the CMB signal and instrumental noise.
The value of $\alpha$ which minimises $\chi^2$ and its formal error are
given by the following equations:

\begin{equation}
E[\alpha] = \frac{\sum_{ij}C_{TM}(\theta_i) {\mbox{\bf $\pi$}}^{-1}_{ij}
			C_{MM}(\theta_j) }
{\sum_{ij}C_{MM}(\theta_i) {\mbox{\bf $\pi$}}^{-1}_{ij}
			C_{MM}(\theta_j)}
\label{eq:alpha2}
\end{equation}

\begin{equation}
\sigma_{\alpha} = \left[ \sum_{ij} C_{MM}(\theta_i){\mbox{\bf $\pi$}}^{-1}_{ij}
			C_{MM}(\theta_j)  \right]^{-\frac{1}{2}} 
\label{eq:sgalpha2}
\end{equation}

This method assumes that all components which are not correlated to the
template $\vm$ average to zero when computing the cross-correlation.
Foregrounds do not average to zero, and so they are supposed to bias
the estimation of $\alpha$ towards positive values. Therefore, the impact
of foregrounds in this analysis is quite similar to that in the previous
method. We must also note that any mismatch in the zero level of the
map would yield spurious correlation, so it is convenient to make
alternative tests, such as repeating analyses after rotating the templates,
 to check whether any measured cross correlation is actually
associated to our templates or not.

We expect method {\it ii} to be particularly
sensitive to the spatial structure of superclusters of galaxies. 
Indeed, under WMAP's 94 GHz
beam, almost all clusters can be regarded as point-like, so we should expect
$C_{MM}(\theta ) =0$ for $\theta > \theta_{FWHM}$. However, 
superclusters of galaxies 
can subtend up to 40 degrees in the sky, and for them it might be possible
to trace their typical angular extent by the study of $C_{TM}(\theta )$.
For these reasons, we shall use method {\it i} for the search of tSZ 
signal in cluster scales, whereas method {\it ii} will be applied on
supercluster templates. Due to the typical size of superclusters (few
degrees), one can include in the analysis different bands of WMAP, as we
show in the following section. Nevertheless, it is worth to remark that, 
in terms of the frequency dependence of the
tSZ effect, the signal should drop around a 14\% from channel V to channel W.
\\

\section{The WMAP data}

We have  mentioned that, due to angular resolution requirements, initially
 we considered only the W band for our analyses with templates of clusters of 
galaxies. Nevertheless, for the sake of comparison, we included the map 
provided by Tegmark, de Oliveira-Costa \& Hamilton (2003), here
 denoted as T03.  
This map weights the
five bands of WMAP according to their sensitivity at every angular scale. 
The result is a map with
as much angular resolution as the W band and smaller presence of both
noise and foreground contaminants. 
However, the properties of its noise component are complex. In our
analyses, we characterised it in a optimistic and/or in a conservative way: 
either we assumed
that the excess power of the map ($\sigma_{excess}\simeq 66\mu$K) was
due to gaussian white noise\footnote{This gaussian noise was assumed to 
follow the spatial pattern of the noise templates provided by the WMAP team}, 
or we simply assigned WMAP's 94GHz band noise pattern to T03.

In order to assess the impact of foregrounds on our results, we performed
our analyses using two different foreground masks: the Kp0 mask and the Kp2
mask, both provided by the WMAP team. These masks remove from the analysis 
those pixels showing temperature fluctuations that trespass a given threshold 
computed from the histogram of the K-band data. The digit in the name of the
mask is proportional to the threshold, so Kp0 is twice as
conservative/strict as Kp2. In these analyses, the HEALPix\footnote{HEALPix URL site: http://www.eso.org/science/healpix/} resolution parameter was $N_{side}=$512.

The relatively large typical size of superclusters allowed us to test the
consistency of our results by including the V band in our analyses. 
Furthermore, {\it clean} maps (in the sense of foreground-free)
provided by the WMAP team (here denoted as {\it ilc} for {\it internal linear
combination} map) and by Tegmark, de Oliveira-Costa \& Hamilton (2003)  were also considered.
 The {\it ilc} map has a typical angular resolution of one degree, whereas T03
has as much angular resolution as the W map. However, at low angular 
resolution, both maps are very similar, so we only retained T03 in our 
analyses. For our purposes of studying superclusters, the resolution of one
degree suffices, so those temperature maps and all supercluster templates 
where pixelized under a coarser resolution, of a pixel size of $\sim 56'$. 
 This angular
degradation allowed us to speed the computation of the cross-correlation
between the WMAP temperature maps and our supercluster templates, and also
made feasible to estimate the covariance matrices of equation 
(\ref{eq:pi}) 
under a reasonable amount of time. We expect that the linear combination
of different bands used to build the T03 map does not remove the tSZ
component of our maps, but leaves it in roughly the same amplitude level.
Our results will justify this hypothesis.\\

\section{Templates}

One crucial step in our procedure will be constructing the templates 
from existing surveys of clusters and superclusters of galaxies, both
in the optical and in the X-range band. We shall
build only tSZ templates, that is, in units of decrements of thermodynamic
temperature. When cross-correlating with WMAP temperature maps, a positive
correlation would point to a detection of tSZ effect. 
In any case, what
matters is the spatial structure of our templates and not their overall
normalisation. 
The conversion into thermodynamic temperature decrements
from X-ray based catalogues will be performed by using existing relations 
between flux and/or luminosity in the X-ray band and antenna
temperature decrements.
For the optical catalogues, further assumptions will be necessary.

In the case of galaxy clusters, there is a fairly large sample of
 catalogues available, both in the X-range and in the optical.
In the northern sky, our analyses will first focus on two X-ray based 
catalogues of clusters of galaxies, i.e., the Northern {\it ROSAT}
All Sky Galaxy (NORAS) Cluster Survey \citep{bohringer00} and
the {\it ROSAT} Brightest Cluster Sample 
(BCS, \citet{ebeling98,ebeling00}). 
The source of these two catalogues is
the published data of the {\it ROSAT} All Sky Survey (RASS, \citet{trumper93},
\citet{vogues99}). The NORAS cluster survey contains 495 sources with
extended X-ray emission, of which 378 are unambiguously identified as
clusters of galaxies. Due to the modest completeness of this catalogue,
(count-rate limit of 0.06 counts/s), we
extend our analysis to the BCS catalogue, which is flux limited and shows
a completeness of around  90\%. This catalogue is built on two different
cluster samples: the first sample \citep{ebeling98} listed
201 clusters of galaxies, whereas the extended sample \citep{ebeling00}
provided 99 new objects. The criteria used in their construction are
not purely X-ray based. Both catalogues (NORAS and BCS) give
luminosities in the 0.1--2.4 keV energy band for every cluster, 
which will be converted into
brightness temperature fluctuations, as it will be shown. 

In the southern hemisphere, \citet{grandi99} extracted from the
RASS data a flux limited sample of
130 clusters of galaxies, with an estimated completeness of around 90 \%.
Redshifts and fluxes in the 0.5 -- 2 keV energy band are given for every
object.  

Analysing {\it ROSAT} Position Sensitive Proportional Counter
pointings, \citet{vikhlinin98} compiled a catalogue of 203 
serendipitously detected galaxy clusters. This catalogue, hereafter denoted
by V98, covers both hemispheres, and therefore complements all catalogues
listed above. 
 V98 provides redshifts and fluxes of members in the energy range of
0.5 -- 2 keV. A conversion into luminosities will be necessary in order
to estimate the antenna temperature fluctuations induced by these clusters.
All catalogues listed so far rely on ROSAT data, 
and hence are all
conditioned by the observation strategy of this experiment.  

For the sake of comparison, we shall include in the analysis
two more catalogues of galaxy clusters.
These remaining two catalogues were built on the basis of existing optical
galaxy cluster surveys, namely APM \citep{dalton97} and ACO
\citep{aco89}. These catalogues provide an homogeneous
sample of galaxy clusters in both hemispheres, and have been processed
after applying selection criteria well different from those used in
X-ray data analysis. However, unlike in the previous cases, these catalogues 
are not directly sensitive to the hot gas causing SZ distortions, and 
assumptions will have to be made when relating optical properties of the
sources (i.e. richness) with antenna temperature decrements. 
Given the angular resolution of the W band, we note that we can safely 
ignore issues related to the gas distribution within cluster scales.

With respect to superclusters, we shall recur to the catalogues provided
by \citet{einasto94,einasto97}. The use of these templates will be essential
when testing the hypothesis of the presence of diffuse hot gas 
distributed in megaparsec scales.

\begin{table}
\caption{SZ cluster sample used to calibrate the template maps}
\begin{tabular}{cccc}
\hline
\hline
Cluster Name &  z & $\Delta T_{SZ}^{RJ} \;^{(a)}$  &$L_X^{(0.1-2.4keV)} \;^{(b)}$ \\
             &    &   (mK) & ($10^{44}~h_{50}^{-2}~erg~s^{-1}$) \\
\hline
A1656(COMA)  &   0.0232  &   $  -0.55 \pm    0.10$	&	  7.26\\
A2256        &   0.0601  &   $  -0.44 \pm     0.09$	&	  7.11\\
A2142        &   0.0899  &   $  -0.90 \pm     0.14$	&	 20.74\\
A478         &   0.09    &   $  -0.92 \pm     0.15$	&	 13.19\\
A1413        &   0.143   &   $  -0.96 \pm     0.11$	&	 13.28\\
A2204        &   0.152   &   $  -0.96 \pm     0.28$	&	 21.25\\
A2218        &   0.171   &   $  -0.75 \pm     0.20$	&	  9.30\\
A665         &   0.182   &   $  -0.91 \pm     0.09$	&	 16.33 \\    
A773         &   0.197   &   $  -0.89 \pm     0.10$	&	 13.08\\
A1835        &   0.252   &   $  -1.34 \pm     0.15$	&	 38.53\\
Z3146        &   0.291   &   $  -0.86 \pm     0.14$	&	 26.47\\
\hline\hline
\end{tabular}
\medskip

(a) Central decrement, from the compilation of \citet{cooray99}.\\
(b) X ray luminosities from the BCS catalogue. 
\label{tab:szdata}
\end{table}

The last step in our template construction procedure was assigning temperature
decrements to the sources in our catalogues. For the case of galaxy clusters in
X-ray based catalogues, it was done in the same way than \citet{cooray99},
 by relating $\Delta T_{tSZ}$ with
$L_X$ from a sample of clusters with measured RJ antenna temperature 
decrements\footnote{It should be noted that quoted values
correspond to the inferred central temperature. 
These numbers are generally derived by fitting 
a $\beta$-profile to the observational data (see individual
references at \citet{cooray99}), so the telescope dependence
is reduced, and a model dependence is introduced.}
Following his discussion, we would expect this relation to have the form
$\Delta T_{tSZ} = a L_x^{b}$. 
A $\chi^2$ minimisation\footnote{We must remark that we are using
here a different statistic than in \citet{cooray99}, so our 
error bars are different.} 
was used to derive the best-fit pair $(a,b)$, using
luminosities in the 0.1-2.4 keV ROSAT band from the BCS catalogue
\begin{equation}
\Delta T_{tSZ}^{RJ} = -(0.24\pm0.06) \Bigg( 
\frac{L_X^{(0.1-2.4keV)}}{10^{44}~h_{50}^{-2}~erg~s^{-1}} 
\Bigg)^{0.47\pm0.09} \; mK  
\end{equation}
The values used for this fit are summarised in table (\ref{tab:szdata}). 
A similar fit can be obtained from the luminosities quoted in
the NORAS catalogue ($a = 0.21\pm0.05 $, $b = 0.53\pm0.09$). 
Clusters extracted from the
RASS-PSC catalogue had fluxes in the same energy range than NORAS and BCS 
(0.1 -- 2.4 keV), so a conversion into flux and $\Delta T_{tSZ}$ was 
straightforward.
For V98 and the southern catalogue of \citet{grandi99}, a previous scaling
of fluxes from the 0.5 -- 2 keV energy band into the 0.1 -- 2.4 keV band 
was first required (we obtained an average conversion factor of
$F_x(0.5-2keV)/F_x(0.1-2.4keV) \approx 0.65$).  
For optically selected clusters (ACO and APM catalogues), 
we have assumed that the observed SZ decrement is proportional 
to some power $n$ of the cluster richness $R$.
Following \citep{Bennett93}, we have probed here their values $n=1,2,3$. 
In addition, we have used the cluster sample described in
table (\ref{tab:szdata}) to calibrate a relation 
with the form $L_X = A (R+1)^{n^{\prime}}$, obtaining 
\begin{equation}
L_X^{(0.1-2.4keV)} = (1.6\pm0.3) (R+1)^{1.4\pm0.2} \; 
10^{44}~h_{50}^{-2}~erg~s^{-1}
\end{equation}
We finally related $L_X$ with $\Delta T_{tSZ}$ by using
the scalings given above. This template will be labelled as
{\it ACO calibrated}.

We conducted two different approaches when modelling the templates of 
superclusters
of galaxies. We first introduced an uniform distribution of hot gas
centred on the optical center of the supercluster and with radius the
angular size of the supercluster, as it is provided by the catalogues.
This template will be referred to as {\it Sph-SC} for 
{\it Spherical Superclusters}, and will probe the presence of hot gas in SC scales.
 However, some correlation present between these structures and
the WMAP CMB maps might be diluted due to the huge angular size ($\theta_{SC}
 \sim 40\degr$) of some members. For this reason, our
 second approach consisted in filling with hot gas a circle of radius one 
degree around every cluster member in each supercluster. In some occasions,
these circles overlap 
along a given direction in those superclusters where member clusters align
forming a filamentary structure. We shall label this template as 
{\it r1dg-SC}. We assumed that 
the spheres of hot gas in both Sph-SC and r1dg-SC templates had the
same density and temperature in every case, and so we assigned identical
tSZ decrements to all of them.

There were two types of sources in our templates, according to their
angular size. Point-like sources refer to all those clusters which
are not resolved by the W channel of WMAP 
\footnote{Some non-resolved clusters are seen in the W map, as A2142.}.
Extended sources refer
to COMA, VIRGO and superclusters. In all cases, the templates have
been convolved with a realistic approximation of the average WMAP 94GHz
beam.
In this initial stage, the pixelization of all templates was such 
that the typical pixel size was around $7'$, ($N_{side}=512$ under
HEALPix pixelization). In the case of superclusters, the templates
were degraded down  to $N_{side}=64$, which corresponds to a pixel size of
$\sim 54'$. Therefore, for superclusters, templates do not contain
$3,145,728$ pixels, but just $49,152$ pixels.\\

\section{Results}

\subsection{Cluster templates}

In the present section we search for correlations between our cluster
templates and WMAP CMB scans. After building the cluster templates,
they were convolved with the symmetrised window functions corresponding to
each of the four differencing assemblies (hereafter DA's) present in the 
W-band. These convolved templates were weighted according to the noise pattern
of each DA, yielding a final template which was produced in exactly the
same way as the averaged CMB map corresponding to the W-band of WMAP.
In order to investigate {\it which} clusters contributed to the CMB
maps, we applied three different amplitude masks, so
only pixels {\it brighter} than some threshold would be used when
searching for correlation. In practice, it consisted in computing the
histograms of pixel amplitudes for each template. These histograms typically
showed a symmetric distribution around zero amplitude (due to the numerical noise introduced by the convolution) plus a bright 
tail, formed by cluster pixels. By increasing/lowering the 
threshold amplitude we merely change the number of pixels in the tail
considered in the analyses. For the three masks of increasing amplitude, 
hereafter labelled as $\nu_1$, $\nu_2$ \& $\nu_3$, we 
selected a sample of 2000, 500 and 50 pixels, respectively.

We used four different CMB maps in this section. One of them was the
W band of WMAP. This channel contains a white noise of an average amplitude
of 175$\mu $K, which makes it significantly more noisy than the map
provided by Tegmark, de Oliveira-Costa \& Hamilton (2003). For comparison,
we included this map (T03) in our analyses as well.
Furthermore, in order to check for systematic
effects in our approach, we repeated the analyses after rotating both maps
5 degrees around the axis orthogonal to the Galactic plane, i.e.,
$\delta l = 5\degr$. The rotated maps will be labelled as {\it R-W} and
{\it R-T03} for W and T03, respectively.
By comparing results from W and T03 to R-W and R-T03, we can {\it i)} check how
our error bars compare with the case of lack of correlation, and {\it ii)},
check that any detected correlation vanishes in the rotated map, so that
we can assign such correlation to sources of angular size $ < 5\degr$. As
mentioned earlier, we try to track the impact of foregrounds by performing our 
analyses in both Kp0 and Kp2 masks.

All results are shown in table 
(\ref{tab:rescl}), for the Kp0 mask.
In this table we define $\beta$ as $\beta \equiv \alpha
 \langle \vm  \rangle$, where $\vm$ denotes the template.
That is, $\beta $ is given in temperature units ($\mu K$), 
and contains the sign of the cross-correlation, i.e., negative for tSZ.
 From this table, one can deduce
that three catalogues are giving some significant tSZ induced signal, namely
BCS, NORAS and de Grandi. For all of them, the statistical significance 
of the detection is higher under the column T03\footnote{Actually,
no detection is reported for deGrandi in the W band. Nevertheless, if
one takes into account the error associated to noise in this band, 
it turns out that the values of $\beta$ under W are compatible (always below
the 2$\sigma$ level) to those quoted in the T03 column.}. This is partially
associated to
the fact that in this case, we are taking the optimistic values of the
noise level for the T03 map. The pessimistic approach would consist in
quoting the values of $\beta$ under the T03 column but with the error
bars given in the W column. We must remark, however, that the error
bars quoted in the optimistic case are not far from (but rather well within)
the null (i.e., free of tSZ emission) values of $\beta$ quoted under the
R-T03 column. 
With respect to optical catalogues, 
we can appreciate that the APM catalogue at 
$n=2,3$ in T03 gives values for $\beta$ which are consistently increasing with
increasing $\nu$, surpassing the threshold of 2$\sigma$ in $n=3$, $\nu_1$.
This trend might be clarified with the second year WMAP data release, for which
noise amplitude is expected to be a factor of $\sqrt{2}$ lower.
In all cases where we claim detection, $\Delta T_{tSZ}$  increases
with the threshold, although its statistical significance diminishes as a
consequence of the smaller amount of pixels. The only slight increase of 
tSZ amplitude with the threshold 
assures that the correlation is raised not exclusively by the
very brightest clusters, but by most of the clusters considered in
each catalogue. Otherwise, the signal would be much more diluted
with decreasing $\nu$\footnote{If only pixels at $\nu_3$ are responsible
for the correlation, then one would expect $\beta$ to drop by a factor of
10, 40 for $\nu_2$, $\nu_1$, respectively.}.
Note that these amplitudes give the average contribution of 
the tSZ effect to WMAP maps, and have to be corrected for the beam and pixel
smearing in order to give the actual cluster decrements.
These analyses were repeated with zero-mean CMB maps, yielding similar
results. The presence of possible offsets in the CMB maps therefore has
negligible impact on our conclusions.

\begin{table*}
 
\begin{tabular}{||c|c|c|c|c|c||} \hline\hline
\multicolumn{6}{c}{\phantom{xxxxxxxxxxxxxxxxx} $\beta \pm \sigma_{\beta}$, 
($\mu$K), [Kp0]} \\
\hline
Template Catalogue & &
W & R-W & T03 & R-T03 \\
  & $\nu_1$ &    20.34 $\pm$     6.83 &    18.57 $\pm$     6.84 &    -1.32 $\pm$     4.75 &     5.21 $\pm$     4.82\\ \cline{2-6}
ACO ($L_X = L_X( R+1)$)  & $\nu_2$ &    17.21 $\pm$    11.26 &    13.23 $\pm$    11.32 &    -2.40 $\pm$     7.60 &     2.72 $\pm$     7.64\\ \cline
{2-6}
  & $\nu_3$ &    32.35 $\pm$    28.69 &    12.46 $\pm$    28.54 &    24.36 $\pm$    18.10 &    -1.49 $\pm$    18.27\\ \hline
  & $\nu_1$ &    13.44 $\pm$     6.65 &    13.89 $\pm$     6.64 &     0.54 $\pm$     3.78 &    -0.35 $\pm$     3.84\\ \cline{2-6}
ACO (n=1)  & $\nu_2$ &    16.22 $\pm$    10.80 &    11.31 $\pm$    10.85 &    -3.41 $\pm$     7.03 &    -0.08 $\pm$     7.14\\ \cline{2-6}
  & $\nu_3$ &    20.69 $\pm$    28.13 &   -23.49 $\pm$    27.96 &     5.36 $\pm$    17.29 &   -35.93 $\pm$    17.56\\ \hline
  & $\nu_1$ &     6.93 $\pm$     6.02 &     9.83 $\pm$     6.05 &    -0.54 $\pm$     3.13 &    -2.72 $\pm$     3.22\\ \cline{2-6}
ACO (n=2)  & $\nu_2$ &     2.76 $\pm$    10.91 &     4.72 $\pm$    11.07 &    -6.81 $\pm$     6.27 &    -3.99 $\pm$     6.54\\ \cline{2-6}
  & $\nu_3$ &    16.47 $\pm$    28.41 &   -29.00 $\pm$    28.50 &     5.19 $\pm$    17.36 &   -30.98 $\pm$    18.00\\ \hline
  & $\nu_1$ &    -0.07 $\pm$     4.79 &     4.36 $\pm$     4.91 &    -2.40 $\pm$     2.26 &    -2.26 $\pm$     2.41\\ \cline{2-6}
ACO (n=3)  & $\nu_2$ &     0.66 $\pm$    10.23 &     2.34 $\pm$    10.44 &    -3.81 $\pm$     5.53 &    -3.33 $\pm$     5.91\\ \cline{2-6}
  & $\nu_3$ &   -24.07 $\pm$    30.24 &   -34.18 $\pm$    30.92 &   -22.77 $\pm$    17.74 &   -10.30 $\pm$    19.30\\ \hline
  & $\nu_1$ &    -3.37 $\pm$     8.94 &    -2.22 $\pm$     8.93 &    -5.67 $\pm$     4.16 &    -3.89 $\pm$     4.15\\ \cline{2-6}
APM (n=1)  & $\nu_2$ &    -2.19 $\pm$    15.96 &     6.15 $\pm$    15.95 &    -7.57 $\pm$    10.40 &   -11.41 $\pm$    10.42\\ \cline{2-6}
  & $\nu_3$ &     3.77 $\pm$    31.99 &    29.28 $\pm$    32.05 &    -3.05 $\pm$    21.92 &     2.82 $\pm$    22.09\\ \hline
  & $\nu_1$ &    -5.58 $\pm$     5.74 &     1.57 $\pm$     5.76 &    -3.24 $\pm$     2.55 &    -3.69 $\pm$     2.58\\ \cline{2-6}
APM (n=2)  & $\nu_2$ &   -10.63 $\pm$    12.45 &     9.96 $\pm$    12.45 &    -9.62 $\pm$     6.27 &    -9.10 $\pm$     6.30\\ \cline{2-6}
  & $\nu_3$ &    -0.54 $\pm$    31.73 &    36.01 $\pm$    31.78 &   -13.10 $\pm$    19.49 &    -3.33 $\pm$    19.95\\ \hline
  & $\nu_1$ &    -4.16 $\pm$     3.39 &    -0.01 $\pm$     3.43 &    -3.52 $\pm$     1.48 &    -1.50 $\pm$     1.52\\ \cline{2-6}
APM (n=3)  & $\nu_2$ &    -6.62 $\pm$     9.07 &     5.25 $\pm$     9.17 &    -5.93 $\pm$     4.10 &    -7.19 $\pm$     4.22\\ \cline{2-6}
  & $\nu_3$ &   -20.71 $\pm$    30.95 &    30.18 $\pm$    31.09 &   -17.91 $\pm$    18.20 &    -8.31 $\pm$    18.84\\ \hline
  & $\nu_1$ &    -3.75 $\pm$     3.05 &    -0.32 $\pm$     2.98 &    -7.24 $\pm$     1.32 &    -0.14 $\pm$     1.29\\ \cline{2-6}
de Grandi  & $\nu_2$ &     2.10 $\pm$    10.45 &     2.75 $\pm$    10.27 &   -21.49 $\pm$     5.19 &     5.31 $\pm$     5.01\\ \cline{2-6}
  & $\nu_3$ &     6.37 $\pm$    28.68 &   -27.48 $\pm$    28.47 &   -48.58 $\pm$    17.72 &    25.55 $\pm$    17.41\\ \hline
  & $\nu_1$ &   -18.10 $\pm$     5.30 &     6.61 $\pm$     5.37 &    -8.44 $\pm$     2.50 &    -2.76 $\pm$     2.57\\ \cline{2-6}
BCS  & $\nu_2$ &   -29.87 $\pm$    11.40 &    16.66 $\pm$    11.62 &   -19.00 $\pm$     6.63 &    -8.25 $\pm$     6.81\\ \cline{2-6}
  & $\nu_3$ &   -64.06 $\pm$    31.78 &    19.59 $\pm$    31.96 &   -39.30 $\pm$    18.43 &   -12.30 $\pm$    18.62\\ \hline
  & $\nu_1$ &   -10.95 $\pm$     5.47 &     8.12 $\pm$     5.44 &   -10.47 $\pm$     2.64 &     4.12 $\pm$     2.69\\ \cline{2-6}
NORAS  & $\nu_2$ &   -15.04 $\pm$    11.06 &    18.17 $\pm$    10.97 &   -17.89 $\pm$     6.50 &     3.79 $\pm$     6.60\\ \cline{2-6}
  & $\nu_3$ &   -29.65 $\pm$    29.16 &    42.09 $\pm$    28.51 &   -49.54 $\pm$    17.30 &     1.50 $\pm$    17.37\\ \hline
  & $\nu_1$ &     5.52 $\pm$     4.31 &     2.98 $\pm$     3.92 &     0.24
$\pm$     1.89 &    -0.06 $\pm$     1.73\\ \cline{2-6}
V98  & $\nu_2$ &    12.94 $\pm$    12.20 &    11.11 $\pm$    11.22 &    -0.56 $\pm$     6.68 &     1.18 $\pm$     6.11\\ \cline{2-6}
  & $\nu_3$ &    17.87 $\pm$    33.01 &    26.42 $\pm$    29.55 &    -1.32 $\pm$    19.89 &    -5.13 $\pm$    17.75\\ \hline
  & $\nu_1$ &     0.50 $\pm$     2.44 &     1.12 $\pm$     2.37 &     0.34 $\pm$     1.07 &     0.19 $\pm$     1.05\\ \cline{2-6}
Vogues  & $\nu_2$ &    14.91 $\pm$     8.78 &    11.10 $\pm$     8.61 &     3.47 $\pm$     4.07 &     0.86 $\pm$     4.05\\ \cline{2-6}
  & $\nu_3$ &    45.18 $\pm$    31.03 &    37.81 $\pm$    30.22 &    10.09 $\pm$    18.51 &     1.57 $\pm$    18.23\\ \hline
\hline

\end{tabular}

\medskip
\caption[tab:rescl]{
This table shows the results of method {\it i} on
cluster templates for the mask Kp0 for the {\bf real} beam.
$\beta$ is defined from $\alpha$ as $\beta \equiv \alpha \langle \vm \rangle$,
it is given in $\mu K$ and contains the sign of the cross-correlation, (i.e.,
$\beta <0$ in the presence of tSZ induced signal).
 Similar results are found for mask
Kp2. Each template has been cross-correlated to the W band of WMAP, to the
combined map of WMAP produced by Tegmark, de Oliveira-Costa \& Hamilton (2003), T03, and to both maps
rotated by $5\degr$ in galactic longitude, R-W \& R-T03, (see columns).
For each
template, we show in different rows results for different thresholds applied
on the templates. More severe threshold (more luminous clusters) give higher
amplitudes for the mean $\Delta T_{tSZ}$ in those catalogues
where a significant
detection is reported, namely, BCS, NORAS and de Grandi, although the
latter shows significant detection only in the T03 map. Note that the
{\it optimistic} estimates error bars for T03 are not too far from the
null detections listed in the R-T03 column.}

\label{tab:rescl}

\end{table*}

\subsection{Supercluster templates}

\begin{figure}
\plotone{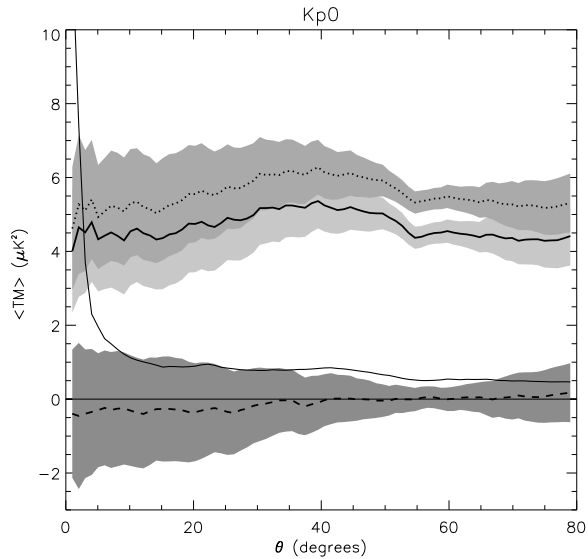}
\caption{This figure shows the cross correlation function 
$\langle T M \rangle$ computed from our r1dg-SC template and the V (thick
solid line), W (dotted line) and T03 (thick dashed line) CMB maps. Each
line is centred in a shaded region limiting the 1-$\sigma$ confidence region
computed from Monte Carlo realizations.
We associate the high amplitude of the cross correlation for the V \& W bands
to foregrounds. T03 gives results compatible with zero. The auto correlation
function $\langle M M \rangle$ amplified by a factor of 10 is displayed by 
the thin solid line. Analyses under the Kp2 mask yielded very similar results.
}
\label{fig:spcl}
\end{figure}

We next describe the results obtained after applying our method {\it ii}
on the V, W \& T03 maps. We restricted our analyses of 
correlation functions to a maximum angle of $\theta_{max} = 80\degr$, as 
this is roughly twice the largest size of the sources in our catalogues.
In figure (\ref{fig:spcl}) we show the cross correlation functions for
Kp0 and r1dg-SC. 
Shaded regions are given by the Monte Carlo simulations and 
show the uncertainty region around $\langle T M\rangle$ associated to 
instrumental noise and cosmic variance. For the T03 case, we used
the {\it optimistic approach} to characterise the noise field. From the
similarity of all shaded areas in our three bands, we conclude
that instrumental noise cannot be of importance at the angular scales
probed by our supercluster template.
We can see that for the V \& W bands, 
the r1dg-SC template gives a positive correlation with the CMB 
scans. Both bands give roughly the same amplitude for the cross-correlation,
and are also very close to those obtained under the Kp2 mask.
However, if the template is rotated 45 degrees in the direction of galactic
longitude, the cross-correlation does not drop, but remains in roughly
similar levels. 
Testing the possibility of a simple offset in the zero-level between the 
V \& W bands and T03, this analysis was repeated 
{\it after subtracting the mean of the Supercluster templates}. In this case,
the cross correlation function was compatible with zero for all angles.
 Moreover, this result is also obtained after applying our method {\it ii}
on the T03 map: the cross-correlation function is not significantly different
from zero in the $\theta$ range of $[0\degr,80\degr ]$\footnote{At $\theta
\simeq 60\degr$ we found an excess of anticorrelation which dissapeared after
removing the quadrupole in the CMB temperature map.}
We know that the T03 map is supposed to be cleaned from foreground signal
to a much greater extent than the V \& W bands. And we have also seen in our
analysis of clusters that the tSZ signal was not removed when combining the
different frequencies to build the T03 map, at least in the cluster
scales. For the large scales (low multipoles) we have also checked from 
figure (6) of Tegmark, de Oliveira-Costa \& Hamilton (2003) that the tSZ 
signal is preserved. Therefore, we associate
the correlation present in the V \& W bands to either an offset or 
a diffuse foreground component
present in these bands but mostly absent in the T03 map. Accordingly, we shall
try to give physical interpretation to those results obtained 
from the T03 map exclusively.

\begin{figure}
\plotone{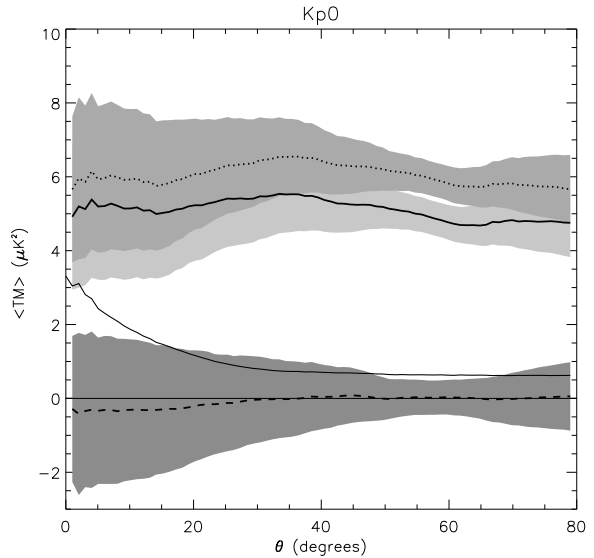}
\caption{Same as in Fig.(\ref{fig:spcl}), but for the Sph-SC template.
}
\label{fig:spcl2}
\end{figure}

In the T03 map, the Sph-SC catalogue gave very similar results 
to those of r1dg-SC, (see figure (\ref{fig:spcl2})).
Our statistical analysis (method {\it ii}) failed to report any
significant detection in any case, but allowed us to limit the comptonization
parameter associated to these angular scales to be $y < 1.86\times 10^{-8},
2.18 \times 10^{-8}$ for 
Sph-SC under the Kp0, Kp2 masks, respectively, at 95\% c.l.
 The r1dg-SC template yielded $y < 8.59\times 10^{-9} $ for both masks, 
and again at 95\% c.l. 
However, one must keep in mind that our modelling of
the Supercluster structure has been rather simplistic, and that these limits
do not observe the inaccuracies of our model when describing the density
and temperature distribution of the diffuse gas.

\subsection{Ring analysis}

In previous subsections, we have demonstrated that
there is evidence of tSZ-induced signal detection using X-ray cluster
templates, while we do not have such detection either for
optical cluster templates or for supercluster catalogues. 
In order to check the robustness of these numbers, we
have also performed a ``ring'' analysis, similar to that utilized
previously in \citet{Bennett93} and \citet{banday96}.
We then compute the difference between the temperature 
in the pixel where we expect to have a cluster, and 
the mean temperature in a ring of radius between $\theta_1$ and
$\theta_2$. We used here $\theta_1=$1.5FWHM, and
$\theta_2=\sqrt{2}\theta_1$.
With this analysis,  we should reproduce
similar amplitudes for the above quoted detections, with the advantage that
now we are particularly unsensitive to a possible zero offset of data. 
The obtained values using the Kp0 mask and the W \& T03 maps are 
shown in table \ref{tab:rings}. Error bars take into account the
temperature correlation of the CMB component among different pixels.
From this table, we confirm the detection of BCS, NORAS and
de Grandi (this one only in the T03 map)  
catalogues. The obtained numbers are comparable to 
those values of $\beta$ using $\nu_2$ (which roughly corresponds 
to use the central pixel of all clusters in each catalogue), and 
the signal dissappears when analysing the rotated maps (R-W and R-T03).  
The other catalogues (V98, Vogues, ACO and APM) provide no detection,
although we marginally see the aforementioned trend in APM using T03.

In order to perform a similar test to our study on supercluster
scales,  we have applied a ring analysis in which we substract 
the mean temperature in annuli of inner/outer radii [$\theta_1=$1.5FWHM,
 $\theta_2=1\degr$] from the mean temperature in annuli limited
by the inner/outer radii [$\theta_2$, $\theta_3=\sqrt{2}\theta_2$], 
in all pixels where we have a cluster member of the supercluster.
As for our method {\it ii}, we find no trace of tSZ emission due to
diffuse gas in \citet{einasto94,einasto97} catalogues, nor in cluster
or superclusters scales.

Before ending this subsection, we would like to remark the agreement 
between methods {\it i} and {\it ii} and the ring analysis.

\begin{table}
\caption{Results of the ring analysis on cluster templates for the
W and T03 maps, using the Kp0 mask. We also include the
results derived rotating the maps $5\degr$ (R-W and R-T03). 
All values are in $\mu K$. }
\begin{tabular}{ccccc}
\hline
\hline
Catalog & W & R-W & T03 & R-T03 \\
\hline
\hline
BCS    & $-28.1\pm11.0$   & $-5.6\pm10.9$ & $-15.3\pm5.0$  & $+0.3\pm4.9$ \\
NORAS  & $-16.6\pm8.6$    & $+4.5\pm8.6$  & $-9.9\pm3.9$  & $-0.7\pm3.9$ \\
Grandi & $-2.4\pm16.2$    & $-1.0\pm16.0$ & $-27.3\pm7.4$  & $+4.5\pm7.3$ \\
V98    & $+4.6\pm12.6$    & $+2.2\pm12.5$ & $-0.9\pm5.7$  & $+4.6\pm5.7$ \\
Vogues & $-6.3\pm13.3$    & $-7.4\pm13.3$ & $-11.2\pm6.1$  & $-7.7\pm6.1$ \\
\hline
ACO    & $+4.3\pm2.7$     & $+5.2\pm2.7$  & $-0.3\pm1.2$  & $+0.3\pm1.2$ \\
APM    & $-5.1\pm5.6$     & $+3.0\pm5.6$  & $-3.5\pm2.6$  & $-0.2\pm2.6$ \\
\hline
\hline
\end{tabular}
\medskip

\label{tab:rings}
\end{table}

\section{Discussion and Conclusions}

We have used optical (Einasto, APM, ACO) and X-ray based (BCS, NORAS,
de Grandi, V98, Vogues) catalogues to locate in the sky tSZ emitting 
sources as clusters and superclusters of galaxies. Using scaling relations
present in the literature, we have assigned microwave fluxes to our
sources, and built tSZ catalogues as they would be seen by the W band of
WMAP. We have cross correlated these catalogues with WMAP using two different
statistical approaches for clusters and superclusters. For the BCS,
NORAS and de Grandi cluster catalogues, we claim detections at the
2-5 $\sigma$ level, (except for deGrandi, which gives no detection in the
W band). In these cases, we have checked that most of the clusters
(and not only the brightest) are contributing to the correlation, with 
amplitudes of typically 20-30 $\mu$K in the WMAP scans. 
We fail to detect any statistically significant signal from the (optical)
 ACO and APM catalogues, (although the latter has chances to give a
detection if the noise level decreases after the second year of observations).
 This may reflect the fact that optical catalogues
are not sensitive to the temperature of the IGM in clusters, and might
include many clusters that, being present in the optical, contribute with
negligible distortions to the tSZ signal, hence diluting the overall
cross-correlation.
We believe the impact of foregrounds to be of
negligible relevance in our results, provided the fact that our method
yields very similar results for different masks (Kp0 and Kp2) and
different CMB scans, (W band of WMAP and the clean map of Tegmark, de
Oliveira-Costa \& Hamilton (2003,T03)). Moreover, all these results are
recovered when a ring analysis (consisting in studying the temperature
fluctuations in annuli centered on cluster pixels) is applied on our 
catalogues.

With respect to our supercluster templates, we have tested the hypothesis
of hot diffuse gas comptonizing the CMB photons in megaparsec scales.
We have built two different supercluster templates, depending upon we
place the hot gas uniformly in the superclusters or concentrated
around the cluster members. By computing the cross correlation function
between these templates and the CMB scans, and comparing it with the
template auto-correlation function, we have traced the level at which
supercluster induced signal is present in the CMB maps. We find that results
depend largely on the CMB map we cross-correlate with: the V \& W WMAP
bands give positive cross-correlation, several $\sigma$-levels above
zero. This correlated signal does not vanish when the template is rotated
45 deg in galactic longitude, but drops to zero in the T03 map.
The T03 map keeps the same zero temperature level of all bands of WMAP, 
but combines them in a multipole dependent manner to minimize the presence of
foregrounds. Hence, we point to an offset or 
foregrounds as the most likely responsibles
for the different cross-correlation present between our supercluster
templates and the W \& T03 CMB maps, and we discard the V \& W bands from
our cross-correlation function analyses.  When applying our method {\it ii}
on T03, we find no significant cross-correlation\footnote{The ring analysis
applied on supercluster catalogues of \citet{einasto94, einasto97} also fails
to detect any tSZ induced signal.}, and this allows us to
place the following constraint on the {\it supercluster induced} 
comptonization parameter $y_{SC}$ associated to the Einasto supercluster 
catalogues: $y_{SC} < 2.18 \times 10^{-8}$ at the $2\sigma$ or 95\% confidence level. 
These figures, however, do not
account for the uncertainty related to the modelling of the gas distribution.
Finally, we conclude noting that any use of 
the V \& W bands in cross-correlation analyses is at best compromised 
by the presence of a diffuse foreground component, at least at the level
of a few microkelvin. 

\section*{Acknowledgments} 

We are particularly grateful to A.J.Banday and S.Zaroubi for useful comments
and suggestions. 
The authors acknowledge the financial support provided through the European
Community's Human Potential Programme under contract HPRN-CT-2002-00124, CMBNET.
Some of the results in this paper have been derived using the HEALPix
package, \citep{healpix}. 
We acknowledge the use of the Legacy Archive for Microwave
Background Data Analysis (LAMBDA, http://lambda.gsfc.nasa.gov). 
Support for LAMBDA is provided by the NASA Office of Space Science.
CHM acknowledges the use of computational resources at the Astrophysikalisches
Institut Potsdam, (AIP).\\

\label{lastpage}

\end{document}